\documentstyle[12pt]{article}
\input epsf.tex
\input fps.sty
\def\fpsangle{270}

\newcommand{\amax}{a_{\rm max}}
\newcommand{\phib}{\bar{\phi}}
\newcommand{\vb}{\bar{V}}
\newcommand{\gbar}{\bar{g}}
\newcommand{\rhob}{\bar{\rho}}
\newcommand{\nperp}{N_{\perp}}
\newcommand{\Ccal}{{\cal C}}

\newcommand{\be}{\begin{equation}}
\newcommand{\ee}{\end{equation}}
\newcommand{\bal}{\begin{array}{l}}
\newcommand{\eal}{\end{array}}
\newcommand{\bea}{\begin{eqnarray}}
\newcommand{\eea}{\end{eqnarray}}
\newcommand{\pat}{\partial}

\newcommand{\lxp}{\lambda x^+}
\newcommand{\xp}{x^+}
\newcommand{\lxm}{\lambda x^-}
\newcommand{\xm}{x^-}

\newcommand{\al}{\alpha}

\newcommand{\dpds}{\frac{d\phi}{ds}}

\newcommand{\greapp}{\stackrel{>}{\sim}}
\newcommand{\rarr}{\rightarrow}
\newcommand{\half}{\frac{1}{2}}
\newcommand{\ac}{\bar{a}}
\newcommand{\ec}{\varepsilon}
\newcommand{\scal}{{\cal S}}

\def\epsilon{{\delta M}}
\def\0{P}

\def\xp{x^+}
\def\xm{x^-}

\begin{document}
\baselineskip 16pt

\begin{titlepage}

\begin{flushright}
MIT-CTP-2529 \\
CALT-68-2056 \\ 
gr-qc/9604058 
\end{flushright}

\vskip 0.1truecm

\begin{center}
{\large {\bf Quantum Gravity and \\ 
Turning Points in the Semiclassical Approximation$^*$}} 
\end{center}

\vskip 0.6cm

\begin{center}
{\bf Esko Keski-Vakkuri}
\vskip 0.2cm
{\it California Institute of Technology \\
     Pasadena CA 91125, USA. \\
     e-mail: esko@theory.caltech.edu }
\vskip 0.2cm
and
\vskip 0.2cm
{\bf Samir D. Mathur}
\vskip 0.2cm
{\it Center for Theoretical Physics \\
    Massachusetts Institute of Technology \\
    Cambridge MA 02139, USA.\\
    e-mail: me@ctpdown.mit.edu }
\end{center}
\vskip 0.4cm
\begin{center}
{\em (To appear in Phys. Rev. D) }
\end{center}
\vskip 0.4cm

\begin{center}
{\small {\bf Abstract:}}
\end{center}

\noindent
{\small The wavefunctional in quantum gravity 
gives an amplitude for 3-geometries and
matter fields. The four-space is usually 
recovered in a semiclassical approximation where
the gravity variables are taken to oscillate 
rapidly compared to matter variables;
this recovers the Schr\"{o}dinger evolution for the matter. 
We examine turning points in the
gravity variables where this approximation appears to be troublesome. 
We investigate
the effect of such a turning point on the matter wavefunction, in simple
quantum mechanical models and in a closed minisuperspace cosmology.
We find that after evolving sufficiently far from
the turning point the matter wavefunction recovers to a form close to that
predicted by the semiclassical approximation, and we compute the 
leading correction (from `backreaction') in a simple model.
We also show how turning points can appear in the gravitational sector
in dilaton gravity. We give some remarks on the behavior of the 
wavefunctional in the vicinity of turning points in the
context of dilaton gravity black holes.
}
\rm
\noindent
\vskip 1 cm

\begin{flushleft}
$^*$ {\small This work was supported
in part by funds provided by the U.S. Department
of Energy under cooperative agreement DE-FC02-94ER40818 and 
by a grant DE-FG03-92-ER40701.}
\end{flushleft} 

\end{titlepage}

\newpage

\section{Introduction}

An interesting issue in formulations
of quantum gravity is to study how the long-distance physics is recovered at
the semiclassical limit. 
If one is interested 
in asking questions about unitarity, one could take
the point of view that it is natural to consider
 wavefunctions describing states; thus one might like to 
consider the wavefunctional
giving amplitudes for 3-geometries and matter configurations on 
these 3-geometries.
The `evolution' of this wavefunctional is given by the 
Wheeler--DeWitt (WDW) equation. In this framework, it has been
shown \cite{SC} how the usual notion of 
space-time is recovered in a semiclassical
approximation of the above wavefunction. 
One takes the gravity variables
to be `rapidly oscillating' while the matter variables 
are `slowly varying'. A WKB
approximation to the gravity wavefunction recovers the Schr\"{o}dinger 
equation for the matter
wavefunction propagating in a WKB `time', which we identify as the 
fourth co-ordinate needed
to describe spacetime.

This scheme would appear to be problematic when the gravity variables encounter
a turning point, where they cannot be rapidly oscillating any more. 
What  will be the effect on the matter wavefunction if the gravity 
variables pass
through such a turning point? This is the question that we investigate
in this paper.

We begin this investigation first with the simple quantum mechanical analogue
of a light particle (`matter') coupled to a heavy particle
(`gravity').  We   find two basic  cases:

a) Let $\Delta t$ be the `time interval' around the turning 
point for which the heavy mode 
is oscillating slower than
the light mode. Suppose the light particle wavefunction does not
evolve significantly over the interval $\Delta t$. Then, as we would naturally 
expect, there
is no significant effect on the light particle  
wavefunction of the fact that the
heavy mode passed through the turning point.

b) In the second case, the light particle wavefunction 
evolves significantly
in the period $\Delta t$. Now the light particle wavefunction departs from the 
wavefunction that is
predicted semiclassically, in the interval $\Delta t$. It turns out however 
that after the
heavy particle recedes from the turning point and its wavefunction 
becomes rapidly oscillating
again, the light particle wavefunction again approaches the form that
it would have had if the turning point had been ignored. We compute the leading
order correction arising from the turning point.

Turning points in gravity-matter systems are not unusual\footnote{For example,
the De Sitter minisuperspace model \cite{QC} has a turning point in the
gravitational sector, with a finite tunneling direction. This fact has
been used in the suggestion of tunneling boundary condition as an 
initial boundary condition for the 
Wheeler--DeWitt wavefunction \cite{VIL}. The behavior and uniqueness of
the wavefunction with such an initial boundary condition was studied
in \cite{VV}.}. As a simple example, we discuss a 
minisuperspace toy model of matter in a closed
cosmology. In this case there is a turning point in the gravitational sector
where the expansion of the universe stops and the recollapse begins.
The semiclassical approximation breaks down locally around this turning
point, and we apply our analysis to find how the approximation
is recovered after passing through the turning point. 

Next we show how turning points
 arise in the description of the geometry of black holes. One puzzle 
which has arisen 
in this context is how different choices of foliations of the black hole
spacetime seem to 
affect the evolution of quantum matter. On one hand, it has been
pointed out by several authors that it is possible to foliate the black hole
spacetime with hypersurfaces on which the evolution is regular all through
(the so-called `nice slices')
leading to the loss of unitarity \cite{NS}. On the other 
hand, examples of irregular
behavior and an apparent breakdown of semiclassical 
approximation (even outside the horizon and far away from the singularity) 
with a different choice of foliation have 
been found \cite{samir,klmo}\footnote{For related issues, 
see \cite{lmo}. Other potential issues of quantum gravity in black holes
were discussed in \cite{lo}.}.  This might be closely related
to the appearance of large commutators in the
variables near the horizon in the treatment of \cite{SVV}.
However, it has also been argued that inclusion of
the backreaction of the Hawking radiation restores the validity
of the semiclassical approximation \cite{bpp}. 

We study different 
foliations of the black hole spacetime, and find that 
the set of spacelike hypersurfaces
that occur in this spacetime fall into two categories, which we 
call I and II. The
Wheeler--DeWitt wavefunctional cannot be evolved from
category I to category II hypersurfaces without passing 
through a turning point. The evolution studied in \cite{samir,klmo} takes 
the matter
wavefunction close to such a turning point. This might then explain 
the apparent breakdown of the semiclassical approximation. We anticipate
(although we do not prove this here) that
this turning point is similar to case b) above. By contrast the evolution of a 
particle in Minkowski space  is presumably
of kind a) above. But by the above discussion of case b), we do 
not expect any large effect of the turning point on the matter wavefunction
in the subsequent evolution after the turning point, even in
the black hole geometry. 

In the above language, a 
foliation with `nice slices' corresponds to all
hypersurfaces lying in category II. 
Then  no turning point is encountered, and
the above issue with the semiclassical approximation 
does not arise. 

The plan of this paper is as follows. In section 2, we start with a simple
quantum mechanical analogue of the Wheeler--DeWitt equation. We introduce
a turning point into the model, and discuss the semiclassical approximation
in the presence of a turning point. We identify the reasons for the breakdown
of the approximation in the vicinity of the turning point, and make order
of magnitude estimates for the size of this region. Then, we study if and how
the semiclassical approximation is recovered after passing through a turning
point, by a comparison with an exact solution of the Schr\"{o}dinger equation.
We calculate the leading correction to the semiclassical approximation, due
to the effect of the turning point. 
In section 3, we apply this analysis to a simple minisuperspace model of
matter in a closed cosmology, and show that there are no large violations
of the semiclassical approximation (after the turning point), if the
quantum matter is lighter than Planck mass. In section 4, we discuss 
foliations of black hole spacetime and show how turning points can arise
(in dilaton gravity). We examine the behavior of the 
wavefunction near the turning point and show that the semiclassical
approximation breaks down locally at the turning point. 
Finally, we make some remarks on the potential
tunneling issues at the turning point.

The semiclassical approximation in dilaton gravity and the problem of
time has been also been investigated in \cite{dealwis}, and more recently
in \cite{demkief}.

\section{Simple Quantum Mechanical Examples}

\subsection{A Heavy and a Light Particle}

The simplest  model for a discussion of the semiclassical 
approximation is given by the quantum mechanics of a heavy particle coupled to
a light particle. Let a  heavy particle of mass
$m_1$ move in a potential $U(x_1)$. Let  a light particle of mass $m_2$ 
be coupled to the heavy particle through a potential $u(x_1,x_2)$.
The total Hamiltonian of the system is thus
\be
      H \equiv H_1 + H_{12} = 
    \frac{p^2_1}{2m_1} + U(x_1) + \frac{p^2_2}{2m_2} + u(x_1,x_2) \ .
\label{c21}
\ee
We consider the quantum  Hamiltonian  $\hat{H}$ and study
its zero energy eigenequation
\be 
    \hat{H}  \  \Psi (x_1,x_2) = 0.
\label{c22}
\ee 
This equation can be regarded as a simple analogue to the Wheeler--DeWitt
equation of a gravity-matter system, if the potential $U(x_1)$ is assumed
to be of the form $U(x_1)= m_1V(x_1)$. Then the heavy particle is analogous
to the gravitational sector and the light particle is analogous to the
matter sector in quantum gravity.
We will now review quickly how the semiclassical approximation is applied
to find an approximate solution of (\ref{c22}). The standard discussion
\cite{REW} goes as follows. To find an approximate solution, we
start with an ansatz
$$
  \Psi (x_1,x_2) = e^{\frac{i}{\hbar} 
     [m_1 \scal_0 + \scal_1 + m^{-1}_1 \scal_2 + \cdots ]}
$$
and solve the equation order by order in $m_1$. This procedure yields
at order ${\cal O} (m^0_1)$ an approximate solution 
$\Psi_{s.c.} (x_1,x_2)$ to (\ref{c22}) with the following
properties. The wavefunction $\Psi_{s.c.}$ can be factorized as
\be
        \Psi (x_1 ,x_2) = \psi_{\rm WKB} (x_1) \ \chi (x_1, x_2) \ ,
\ee
where $\psi_{\rm WKB}$ is a WKB wavefunction for the heavy particle,
\be
      \psi_{\rm WKB} (x_1) = 
  \frac{1}{(-2 V(x_1))^{(1/4)}} \ e^{\pm \frac{i}{\hbar} m_1 \scal_0(x_1)} \ ,
\ee
with the exponent $\scal_0$ satisfying the Hamilton-Jacobi equation for the
zero energy classical motion of the heavy particle
\be
     \half (\frac{d\scal_0}{dx_1} )^2 + V(x_1) = 0 \ .  
\label{c22hj}
\ee
Thus,  
\be
     \scal_0 (x_1) = \int_{x_1} \ dx'_1 \sqrt{-2 V(x'_1)} \ .
\label{c22b}
\ee
Since $d\scal_0 / dx_1$ is the velocity of
the heavy particle,  we can find the classical zero energy
trajectory $x_1 (t)$ through
\be
         t(x_1) = \int_{x_1} \frac{dx'_1}{\sqrt{-2 V(x'_1)}} 
\ee
with the pertinent initial conditions
and thus obtain $x_1(t)$. We can think
that the classical motion of the heavy particle
defines a 'clock' measuring time $t$, often called a 'WKB time'.
(We  now see the benefit of taking  the heavy
particle potential  of the form $U=m_1V$: the time $t$ is
independent of the mass of the heavy particle.) The clock is useful in the
following way. The function $\chi (x_1 ,x_2)$
satisfies the equation 
\be
     i \hbar \frac{d\scal_0}{dx_1} \frac{\pat}{\pat x_1} \chi (x_1 ,x_2)
        = \hat{H}_{12} \chi (x_1 ,x_2 ) 
\label{c23}
\ee
which can be written in a more suggestive form using the fact that
$$
\frac{d\scal_0}{dx_1} \frac{\pat}{\pat x_1 } = \frac{\pat}{\pat t} \ .
$$
The equation (\ref{c23}) is now recognized as a 
time dependent Schr\"{o}dinger equation
for the light particle,
\be
      i \hbar \frac{\pat}{\pat t} \ \chi = \hat{H}_{12}(x_1(t),x_2) \ \chi \ .
\label{c23b}
\ee
[Note that the role that $x_1(t)$ plays in this equation   
resembles the situation in adiabatic 
approximation in quantum mechanics, except that the motion of the 
heavy particle does not have to be 'slow'.]
Now let us summarize what was done. We started by looking at a zero
energy time independent Schr\"{o}dinger equation (\ref{c22}). 
We found an approximate solution, which tells us that
the heavy particle motion is essentially classical whereas the light particle
behaves in a quantum mechanical way. 
Although the starting equation
does not give any time evolution for the total state, the heavy and
the light particle are mutually correlated in such a way that the
light particle can be thought to time evolve according to a 
Schr\"{o}dinger equation with the time measured by the position of the
heavy particle. We point out that since the exponent of the WKB
part of the total wavefunction is proportional to $m_1$,
the heavy particle part of the wavefunction is much
more rapidly oscillating compared to  $\chi$  which describes the light
particle.

Finally, we would like to repeat three characteristic properties
\footnote{It should be noted that
these properties 
differ from those of the Born-Oppenheimer approximation} of the semiclassical
solution:
\begin{enumerate}
\item The wavefunction $\chi$ was not assumed to be an energy
eigenstate of $\hat{H}_{12}$. 
\item The heavy particle is described by a WKB wavefunction.
\item The light particle satisfies a time dependent Schr\"{o}dinger
equation with respect to a time {\em defined} by the classical 
motion of the heavy particle. 
\end{enumerate}
To be
more precise, we call this the leading order semiclassical approximation
to make a distinction with a treatment where 
higher order (${\cal O}(m^{-1}_1)$) corrections would be included.

We will now turn our attention to the question what happens to the
semiclassical approximation if the heavy
particle potential $U$ has classical turning points. Discussions
in the literature about this issue seem to be mostly concerned about
tunneling. However, we will not discuss tunneling effects
but rather study the effects of the existence
of one turning point on the validity of the (leading order) semiclassical
approximation. 

\subsection{Local Breakdown of the Semiclassical Approximation in the
Vicinity of a Turning Point}

Since the WKB approximation  for the heavy particle is a part 
of the semiclassical approximation,
the above approximation scheme must break down near a turning 
point of the heavy particle motion.  
The WKB approximation is
known to be applicable as long as the fractional change 
in the de Broglie wavelength $\lambda_{dB}$
of the particle is small over distances which are of the order of the
wavelength itself. Since the de Broglie wave length is inversely
proportional to the momentum of the particle, and by definition the
momentum approaches zero at a classical turning point, in the vicinity
of the turning point $\lambda_{dB}$ grows very quickly and the WKB
approximation is no longer applicable.

However, the breakdown of the semiclassical approximation
can be due to other reasons than the breakdown of WKB.
Let us now incorporate a turning point into the heavy particle sector and then
make simple order of magnitude estimates about the size of the region 
where the semiclassical approximation is problematic. 

In order to avoid tunneling effects, we assume
that the heavy particle has only one classical turning
point, at $x_1 = a$. It is important to keep in mind that the position
of a turning point depends on the available energy of the heavy
particle. 
Let us expand the light particle wavefunction $\chi$ in energy eigenstates
$\chi_n$
$$
     \chi = \sum^{m}_{n=0} \ c_n \ \chi_n (x_1,x_2)
$$
with eigenenergies $E_n(x_1)$. We will assume \footnote{We will 
assume that $x_1$ is sufficiently
close to the turning point so that we can ignore level crossings.
Thus the light particle Hamiltonian is assumed to not change 
significantly in the region where the WKB approximation for the 
heavy particle breaks down.}
 that the typical separation
of the eigenenergies is $\Delta E$ and that the difference of the 
minimum and maximum energy levels is of the
same order as $\Delta E$.
We can include the energy $E_0$ of the reference level $\chi_0$ into the 
normalization of the total energy of the system (in the choice of the zero of 
the total Hamiltonian $\hat{H}$), therefore we can set $E_0 = 0$. 
This is the normalization 
for the zero energy classical motion of the 
heavy particle with a turning point at $x_1=a$.

Let us fix the additive constant permitted in $\scal_0$ and 
rewrite (\ref{c22b}) as
\be
     \scal_0 (x_1) = \int^a_{x_1} dx'_1 \sqrt{-2 V(x'_1)} 
\label{c24}
\ee
so the heavy particle phase factor vanishes at the turning point $a$. 
Now we investigate how rapidly the heavy particle and light particle
phase factors oscillate near the turning point. Recall that the time
dependence in (\ref{c23b}) arises through the classical 
motion $x_1 = x_1(t)$  of the heavy particle. Near the turning point,
the motion of the heavy particle will slow down. 
Thus the parameter $x_1(t)$ is slowly evolving. 
We  assume that the energy levels are sufficiently 
separated, and approximate the time evolution of the light 
particle wavefunction
by an  adiabatic approximation
\be
     \chi (x_1(t),x_2) = \sum^m_{n=0} \ c_n
     \exp \{ -\frac{i}{\hbar} \int^t_0 \  E_n(x_1(t'))dt' \}
            \ \chi_n (a,x_2) \ ,
\label{c24zz}
\ee
where the time is taken to be zero at the turning point: $x_1(0)=a$.
Now, we can change the integration variable from $t$ to $x_1$. Using 
$\dot{x}_1 = d\scal_0/dx_1$ with (\ref{c22hj}) we get
$$
    \chi \approx \sum_n \ \exp \{ -i\varphi_{l,n} (t)\} \chi_n 
$$
where the phase for level $n$ of the light particle is
\be
         \varphi_{l,n} = \frac{1}{\hbar} 
                \  \int^a_{x_1} dx'_1 \frac{ E_n(x'_1)}
                                          {\sqrt{-2 V(x'_1)}} \ . 
\label{c24b}
\ee
Let us compare this with the heavy particle phase factor 
\be
         \varphi_h = \frac{1}{\hbar} \ m_1 
                \  \int^a_{x_1} dx'_1 \sqrt{-2 V(x'_1)} \ . 
\ee
An essential feature of the semiclassical approximation for 
obtaining the wavefunction $\Psi_{sc}$ is that the
heavy particle sector should oscillate more rapidly than the
light particle sector, or 
$\pat \varphi_h / \pat x_1 \gg \pat \varphi_l / \pat x_1 $. This is
equivalent to 
\be
 \Delta E(x_1) \ll -2 m_1 V(x_1) = m_1 (\frac{dx_1}{dt})^2 \ ,
\label{c25}
\ee
where $\Delta E$ is the range of energies contained in the 
light particle wavefunction.
In other words, the semiclassical approximation is applicable in the region
where the kinetic energy of the heavy particle is much larger than the
range of the light particle energies.
Thus, two features of the semiclassical approximation can be violated
in the vicinity of a turning point:
\begin{enumerate}
\item The heavy particle cannot be described by the WKB approximation:
$d\lambda_{dB} /dx_1 \sim 1$.
\item The oscillations in the light particle sector become comparable
to those of the heavy particle sector.
\end{enumerate}

We make now order of magnitude estimates for the sizes of the regions where
the two conditions are violated. Let us assume that the heavy
particle potential is smoothly varying so that we can use a linear
approximation near the turning point:
$$
  U(x_1) \approx m_1 \ V'(a) \ (x_1 -a) \ ,
$$
and assume that $\Delta E(x_1) \approx \Delta E(a)$.  
Then, the WKB approximation breaks down in a region of size
$$
  \Delta x_1 ({\rm WKB}) \sim (\frac{\hbar^2}{m^2_1 V'(a)})^{\frac{1}{3}} \  .
$$
On the other hand, (\ref{c25}) is violated in a region of size
$$
   \Delta x_1 ({\rm s.c.}) \sim \frac{\Delta E}{m_1 V'(a)} \ .
$$
The larger of the two 
regions $\Delta x_1 ({\rm WKB}), \Delta x_1 ({\rm s.c.})$ is
the size of the region where the semiclassical  approximation cannot
be trusted. The latter region becomes larger than the region of the
breakdown of WKB when 
\be
     \Delta E \greapp [(\hbar V'(a))^2 m_1]^{\frac{1}{3}} \ .
\label{delimit}
\ee

An alternative way to think is that  
the region $\Delta x_1 ({\rm s.c.})$ (where the light 
particle wavefunction oscillates faster than the heavy particle
wavefunction) is large
enough to matter only if the semiclassical evolution in this region would
have caused the light particle levels to evolve through a significant phase,
{\em i.e.}
\be
   \varphi_l (a+\Delta x_1 ({\rm s.c.})) - \varphi_l (a)
    \greapp 2\pi \ .
\label{lpcycle}
\ee
It turns out that the condition (\ref{lpcycle}) is equivalent to
(\ref{delimit}). So, to summarize the results of this section:
\begin{enumerate}
\item If the range of the light particle energies is extremely small,
the semiclassical approximation breaks down near a turning point due to
the breakdown of the WKB approximation in the heavy particle sector.
The size of the breakdown region is $\Delta x_1 ({\rm WKB})$.
\item However, the breakdown of the semiclassical approximation can 
happen in an even larger region when either the light particle energies
have a wide range $\Delta E$ or when the slope $U'$ of the heavy particle
potential is very gentle near a turning point. In this case the breakdown
region has size $\Delta x_1 ({\rm s.c.})$.
\end{enumerate}

The above discussion locates the region where the light particle wavefunction
is not well described by the semiclassical 
approximation. We now ask the question: After the heavy 
particle moves away from the turning point and
the semiclassical approximation becomes good again, what is the net
effect that is obtained in its wavefunction due to the passage
through the turning point?

\subsection{Semiclassical Approximation vs. Exact Results}

We  consider a simple example where everything can be calculated exactly.
Thus it is possible to compare the exact wavefunction with the approximate
semiclassical wavefunction and see how good an approximation the latter is.
For this purpose, let us choose specific potentials $U(x_1), u(x_1,x_2)$ and
consider the model
\be  
  H = \frac{p^2_1}{2m_1} + m_1 r x_1 + \frac{p^2_2}{2m_2} +
\frac{\kappa}{2}(x_1 -x_2)^2 \ .
\label{hamilton}
\ee
This system could be thought to describe a heavy particle 
moving on an incline (with $r$
denoting the slope of the incline) and a light particle moving on a
line, connected to the heavy particle by a spring with a spring
constant $\kappa$.
Let us now find an exact solution of the time independent Schr\"{o}dinger
equation
\be
    \hat{H} \ \Psi (x_1,x_2) = E_{\rm total} \ \Psi (x_1,x_2) \ ,
\label{SE}
\ee
where an arbitrary energy $E_{\rm total}$ reflects the freedom in the choice
of a zero energy ({\em i.e.} freedom in adding an overall constant
to the total Hamiltonian). We separate variables
by changing the coordinates $x_1,x_2$
to the center of mass coordinate $X=(m_1x_1 + m_2x_2)/(m_1+m_2)$ and the 
relative separation $y=x_2-x_1$. Now we write the 
eigenfunction  as $\Psi (x_1,x_2) = F(X)G(y)$.
This gives the two equations
\bea
    \{ -\frac{\hbar^2}{2\mu} \frac{\pat^2}{\pat y^2} 
                 + \frac{\kappa}{2} (y - \frac{r\mu}{\kappa})^2 \}
                         \ G(y) = E \ G(y) \\ 
    \{ -\frac{\hbar^2}{2M} \frac{\pat^2}{\pat X^2} + m_1 r X \} \ F(X) 
                   = (E_{\rm total}-E) \ F(X) \ \ .
\label{mark}
\eea
Here $M=(m_1+m_2)$ is the total mass and $\mu=(m_1m_2)/M$ is the 
reduced mass. Since
$m_1 \gg m_2$, we can assume that $M \approx m_1$ and $\mu \approx m_2$.
We will then speak somewhat loosely and call $X \approx x_1$ as the
position of the heavy particle and $y$ as the position of
the light particle. 
The first equation is the simple harmonic oscillator Schr\"{o}dinger
equation with the familiar discrete 
eigenenergies\footnote{Following the convention in Section 2.1 (see the
discussion before the Eqn. (10)), the zero point
energy is included in $E_{\rm total}$.} $E=E_n=n\hbar \sqrt{\kappa /\mu}$ and
eigenfunctions $G(y)=G_n(y)$. Consider now the second equation. We 
introduce the wavenumber 
$$
  k(X,E_n) = \frac{1}{\hbar}\sqrt{2rMm_1(\frac{E_{\rm total}-E_n}{m_1r}-X)}
$$ 
and write the second equation as
$$
    F_n''(X) + k^2(X,E_n)\ F_n(X) = 0  \ .
$$
The heavy particle has a classical turning point at 
\be
           X = \frac{E_{\rm total}-E_n}{m_1 r} \equiv a(E_n) \ .
\label{c26}
\ee
We would like to emphasize that {\em the location of the turning point
depends on the energy of the light particle}. This is 
different from what was seen in the semiclassical approximation, there the
location of the turning point did not depend on what the light
particle was doing. We will show
that this fact may give rise to additional effects which can be missed
in the leading order semiclassical approximation.
The exact solution of the heavy
particle equation (see {\em e.g.} \cite{Merz}) is 
$$
    F_n(X) = \frac{1}{\sqrt{k(X,E_n)}} \ v_{\lambda} (S(X,E_n)) \ , 
$$
where\footnote{We use the notation $S(X,E_n)$ in the context of exact
solutions, and the notation $\scal_0(X)$ in the context of semiclassical
solutions. The former is defined to contain 
the mass and $\hbar$, while the latter is defined not to 
contain $M$ or $\hbar$.} 
\be
  S(X,E_n) \equiv \int^a_X dX' \ k(X',E_n) 
           = \frac{2}{3} \sqrt{\frac{2rm_1M}{\hbar^2}} 
               e^{i\pi} (a(E_n)-X)^{\frac{3}{2}} 
\label{c26b}
\ee
and $v_{\lambda}$ is a function given in a complex integral form
in \cite{Merz} pp. 134-137.
Here the only thing we need to know about $v_{\lambda}$ is that 
there are two independent solutions, corresponding to
$\lambda = 1/6,\ 5/6 $
and that in the region far away from the turning 
point $v_{\lambda}$ reduces to a
superposition of two WKB solutions:
$$
v_{\lambda} \sim A \ e^{i \int^a_X dX' k(X')} + 
                 B \ e^{-i \int^a_X dX' k(X')} \ .
$$
However, near the turning point it behaves like
$$
     v_{\lambda} \sim k^{3\lambda} \ .
$$
Now we take into account the joining condition at the turning point. Since
the heavy particle is deep in the incline, the wavefunction $F$ can
only be exponentially decaying in the region $X>a$. Then we know that
sufficiently far in the classically allowed region $X<a$ the function
$F$ takes the form
$$
     F_n(X) = \frac{1}{\sqrt{k(X,E_n)}} \ \cos [-S(X,E_n)- \phi(a(E_n)) ] 
$$
where $\phi (a(E_n))$ is a constant phase shift due to the turning point.
Its general form is
\be
     \phi (a(E_n)) = \mu_j \ \xi (a(E_n)) 
\label{pshift}
\ee
where $\mu_j$ is the Maslov index of the corresponding classical trajectory
of the heavy particle (crudely speaking, the number of times the particle
travels through the turning point - in this case $\mu_j=1$) and 
$\xi (a(E_n))$ is a number depending on the shape of the potential
and the location $a(E_n)$ of the turning point. In the case of  a linear
potential (and more generally, of a smooth potential which can be approximated
by a linear potential  at every point), $\xi (a(E_n)) = \frac{\pi}{4}$.
Therefore, for a smooth potential the constant phase shift is independent
of the location of the turning point and thus also independent of the
energy of the light particle. 

Thus, in our example the total exact wavefunction reduces to
\be
   \Psi = \sum_n \ c_n \ \frac{1}{\sqrt{k(X,E_n)}} 
      \  \cos [-S(X,E_n) - \frac{\pi}{4} \ ] \ G_n(y) 
\ee
when $x_1$ is far from the turning point. In Appendix A we reanalyze
the problem using solely semiclassical approximation. This yields an
approximate solution
\be
   \Psi_{\rm s.c.} = \sum_n \ c_n \ \frac{1}{\sqrt{k_0(X)}} 
           \ \cos [ \ \frac{1}{\hbar} M \scal_0(X) 
                  + \frac{1}{\hbar} E_n t_{\rm WKB} (X) \ ] \ G_n(y) 
\ee
where 
$$
\left\{
\begin{array}{l}
       \scal_0(X) = \frac{2}{3} \sqrt{2r} [a(0) - X]^{\frac{3}{2}} \\
       a(0) \equiv E_{\rm total}/(Mr)  \ \ \ .
\end{array} \right.
$$
We can see that the exact result reproduces the phase
behavior of the semiclassical solution in 
the WKB region. Far away from the turning
point $\sqrt{a(0) -X} \approx \sqrt{a(E_n) -X}$, so 
$k_0(X) \approx  k(X,E_n)$ (recall that
$M \approx m_1$). In order to compare the oscillation rates,
we expand 
\be
     S(X,E_n) \approx S(X,0) + E_n \ \frac{\pat S}{\pat E_n} (X,0)
            + \frac{1}{2} \ E^2_n  \ \frac{\pat^2 S}{\pat E^2_n} (X,0) 
            + \cdots \  \ \ .
\label{expand}
\ee
The first term $S(X,0) \approx -\frac{1}{\hbar}M\scal_0(X)$. 
How about the higher order terms?
For a generic potential $U(X)=m_1V(X)$, the first derivative of $S$ is
$$
   \frac{\pat S}{\pat E_n} (X,E_n) = -\frac{M}{\hbar^2} 
     \ \int^{a(E_n)}_X \ \frac{dX'}{k(X',a(E_n))} \ . 
$$   
If we compare this with the WKB time
\be
 t_{\rm WKB} = \frac{M}{\hbar} 
  \ \int^{a(0)}_X \ \frac{dX'}{k_0(X')} \ .
\label{wkbtime}
\ee
we see that
\be
     \frac{\pat S}{\pat E_n}(X,0) = -\frac{1}{\hbar}t_{\rm WKB} \ .
\label{sprime}
\ee
This illustrates how the (WKB) time evolution of light particle states 
arises from the heavy particle sector.
In the present example of the linear potential,
all higher derivatives $\pat^m S / \pat E^m_n$
decay to zero. 
Thus, far away from the turning point there are no lasting effects from 
the reflection from the turning point
(other than the constant phase shift by $\pi /2$)
and the semiclassical approximation is again sufficient.

This is {\em not} true in general.
For instance, the second derivative $\pat^2 S / \pat E^2_n$
does not always decay
to zero but can instead approach a
{\em nonzero} value. This means that there is a leftover phase factor
which is of second order in the light particle energy.

We now give an example of a case
where the higher order corrections do not die out. Consider
the potential 
$$V(X) = \left\{ \begin{array}{l} -bX^2 \ , \ X<0 \\
                                   0 \ , \  X \ge 0 \ \ \ \ .
\end{array} \right. 
$$
The light particle energy
dependent turning point is at $X=a(E_n) = -\sqrt{(E_n -E_{\rm total})/(bM)}$ 
(now we must take $E_{\rm total}<0$) and the wavenumber is 
$$
 k(X,a(E_n)) = \frac{M}{\hbar} \ \sqrt{2b\ (X^2-a^2(E_n))} \ .
$$
In the leading order semiclassical approximation, the wavenumber is 
$$
 k_0(X) = \frac{M}{\hbar} \ \sqrt{2b\ (X^2-a^2(0))} 
$$
with $a(0)= -\sqrt{(-E_{\rm total})/(bM)}$. Now we get
\be
 \frac{\pat S}{\pat E_n} (X,E_n) = \frac{1}{\hbar \sqrt{2b}}
\cosh^{-1}(\frac{a(0)}{a(E_n)} \cosh (-\sqrt{2b}
   \ t_{\rm WKB})) \ ,
\label{commentthis}
\ee
with $t_{\rm WKB}$ defined in (\ref{wkbtime}). 
Thus, $\frac{\pat S}{\pat E_n} (X,0)$ 
satisfies (\ref{sprime}).\footnote{In Eqn. 
(\ref{commentthis}) the semiclassical time $t_{\rm WKB}$ is well defined
only for $|\Delta t_{\rm WKB} | > t_* \equiv \frac{1}{\sqrt{2b}} 
\cosh^{-1} (\frac{a(E_n)}{a(0)})$. The 
notion of semiclassical 
time is ill defined in the immediate vicinity of the turning point.} 
However, for the second derivative we get 
$$
    \frac{\pat^2 S}{\pat E^2_n} (X,E_n) = \frac{1}{2\hbar
             \sqrt{2b} (E_n-E_{\rm total})} 
        \ \frac{X}{\sqrt{X^2 - \frac{E_n-E_{\rm total}}{bM}}}
$$
so far away from the turning point it tends to a finite value
$$
 \frac{\pat^2 S}{\pat E^2_n}(-\infty, 0) = \frac{1}{2\hbar
                       \sqrt{2b} (-E_{\rm total})} \ .
$$
If the curvature of the potential well is very small ($b$ small) or
if the turning point approaches $X=0$, this 
residual energy dependence in the phase factor can be quite significant, and
leads to additional interference effects between different energy
eigenstates. This is the 'imprint' in the wavefunction from the
turning point which we mentioned in the introduction.

To extract a result in more general terms 
notice that the last formula can be rewritten as
\be
   \frac{\pat^2 S}{\pat E^2_n} = \frac{1}{\hbar M} 
        \frac{\sqrt{-V''(a(0))}}
             {[V'(a(0))]^2} \ .
\label{c211b}
\ee 

Then we can see that for
a general potential, if the turning point is near a 
local maximum of the potential, then for
the lowest order corrections to the wavefunction arising from 
the turning point  we must use the expression  (\ref{c211b}). 
After the evolution has moved far from the turning point, the 
light particle wavefunctions far before and after the turning point 
are related by
\be
   \chi_{\rm before}~\equiv~\sum_n c^{\rm before}_n\chi_n   
 \label{c211bpone}
\ee
\be
   \chi_{\rm after}~\equiv~\sum_n c^{\rm after}_n\chi_n   
\label{c211bptwo}
\ee 
\be
   c_n^{\rm after}~=~ c^{\rm before}_n e^{-\frac{i}{\hbar}
       E_n \Delta t_{\rm WKB} + i E_n^2 \frac{\partial^2 S}{\partial E_n^2}}
\label{c211bp}
\ee 
where $\Delta t_{\rm WKB}$ 
depends on the times being compared and the constant 
$\frac{\partial^2 S}{\partial E_n^2}$ is determined from  (\ref{c211b}).
 
In the above we have seen the distortion of the 
wavefunction (as compared to the lowest order semiclassical 
approximation) created  
by the fact that the slope of the heavy particle potential 
at the turning point is different for different values of the 
light particle energy. Thus happens whenever the heavy 
particle potential is not exactly linear. The location of the 
turning point varies
nonlinearly with the energy, and the consequence of this is that
the different energy components of the light particle wavefunction 
pick up a phase that is not linear in their energy, and such a phase 
cannot be accounted for by normal   evolution  using  a linear 
Schr\"{o}dinger equation for the light particle.

There is another source of the distortion of the wavefunction, which 
can happen if the phaseshift (\ref{pshift}) is itself different for 
different positions of the heavy particle turning point. A simple example
of this is given by  $V$ which has the form of a finite potential step: 
now $\phi (E_n)$ is not a simple constant phase shift like $\pi / 4$, 
as we show in Appendix B. This  means that different
light particle energy levels in the full wavefunction have different
phase shifts (\ref{pshift}). Since this is a genuine quantum 
mechanical phase shift, it may be hard to reproduce by 
considering higher orders in the semiclassical approximation scheme.

In conclusion, in the simple examples that we have considered, we
found that there may be small corrections which can modify the wavefunction
slightly away from its leading order semiclassical form. It would be 
interesting to extend this analysis to systems where the heavy and 
light particle sectors are coupled in a more non-trivial way, or to 
systems with more degrees of freedom. 

\section{Example in Minisuperspace Quantum Cosmology}

Now that we have examined how turning points  modify the 
the simple phase correlations predicted by the
leading order semiclassical approximation, we
would like to investigate an example in quantum gravity.
We discuss a simplified quantum cosmological toy model of
quantum matter evolving in a closed universe and check whether the
turning point effects found in the previous section would spoil the
semiclassical approximation in this case. It would be quite surprising
if this should happen in the case of a macroscopic 
universe. At least in our simple model there will turn out to be
no large violations of the semiclassical 
approximation if the quantum matter is lighter
than Planck mass.

We start with
$$
   S = S_{\rm grav. + cl. matter} =  \int \ d^4x \sqrt{-g}
          (\frac{R}{16 \pi G} - \rho_m) \ ,
$$
where we will consider $\rho_m$ as a function related to the 
energy density of the pertinent
classical matter and we will specify it later. The first term in the
action is the Einstein-Hilbert action of classical gravity and $G$ is
the Newton's constant.

Using a spherically symmetric ansatz for the metric,
\bea
    ds^2 &=& \sigma^2 [N^2_0(t) dt^2 - a^2_0(t) d\Omega^2_3] \\ \nonumber
         &\equiv& N^2(t) dt^2 - a^2(t) d\Omega^2_3 \ ,
\eea
where $N,a$ have dimensions of length, $t$ is dimensionless and
$\sigma^2 = 2G/3\pi $, the action reduces to 
\bea
    S &=& \int \ dt L \ ,  \\ \nonumber
    L &=& \frac{N}{2\sigma^2} \ \{ a(1-\frac{\dot{a}^2}{N^2}) - 4\pi^2
                                       \rho_m a^3  \} \ .
\eea
If $\rho_m$ would be a (cosmological) constant, this would be the De Sitter
minisuperspace model \cite{QC}.
Instead, we will choose $\rho_m$ to satisfy
$$
     4\pi^2 \rho_m a^4 \equiv C^2 = {\rm constant} 
$$
which means that we interpret $\rho_m$  as the energy 
density of a classical radiation fluid
filling the universe. The constant $C$ has the dimension of length. 
Now the Lagrangian becomes
$$
     L = \frac{N}{2\sigma^2} \  \{ a(1-\frac{\dot{a}^2}{N^2}) -\frac{C^2}{a}\}
$$
and, varying with respect to N, we get a classical equation of motion
$$
     1 + \frac{\dot{a}^2}{N^2} - \frac{C^2}{a^2} = 0 \ .
$$
The model is time reparametrization invariant, which means that we
need to make a gauge choice and choose a particular $N$. We choose 
$$
     N = \sigma \equiv \frac{1}{M} \ , 
$$
the latter notation is useful since $M$ is essentially the Planck
mass, $M \sim m_{\rm Planck}$. Likewise, $\sigma$ is the Planck length
$\sigma \sim l_{\rm Planck}$. The classical equation of motion
corresponds to the dynamics of a closed radiation filled Robertson-Walker
cosmology, with the solution
$$
  a = \sqrt{C^2 - (\frac{t}{M})^2} \ .
$$
We chose the origin of time so that the expansion of the universe
starts at $t = -MC$, it reaches the maximum size $\amax = C$ at $t=0$,
after which it recollapses to zero size at $t=+MC$. If $MC \approx
\amax / l_{\rm Planck} \gg 1$, the universe will expand to a macroscopic
size and has a macroscopic lifetime. 

When the model is quantized, we obtain the Wheeler--DeWitt equation
\be
  \frac{1}{2} \{ \frac{1}{aM^2} (\frac{\pat^2}{\pat a^2} 
    + \frac{\gamma}{a} \frac{\pat}{\pat a}) + M^2 a 
  (\frac{C^2}{a^2} -1) \} \ \Psi (a) = 0 \ ,
\label{minisuper}
\ee
where $0 \leq \gamma \leq 1$ is a free parameter reflecting an
operator ordering ambiguity in the kinetic term. The 
$\gamma$-dependent term is known to be irrelevant in the WKB regime. (We 
provide an argument for this in Appendix C). Thus,
we expect that the value of $\gamma$ will not affect our 
discussion and will assume   $\gamma = 0$ henceforth. We 
rewrite the WDW equation
then as
$$
    \hat{\cal{H}}_{\rm gr+c.m.} \ \Psi (a) = \frac{1}{2} \{ -\frac{1}{aM^2}
        \hat{p}^2 + \frac{M^2}{a} (C^2 -a^2) \} \ \Psi (a) = 0 \ ,
$$
where $\hat{p} = -i\pat / \pat a$.

Let us now include additional matter in the model. We assume that the
energy of the additional matter is much lower than that of the
classical radiation fluid which was already included in
$\hat{\cal H}_{\rm gr+c.m.}$. We will then treat the additional matter quantum
mechanically and consider it to propagate in the background of the
expanding/recollapsing universe. The WDW equation is now
$$
    \{ \hat{\cal H}_{\rm gr.+c.m.} + \hat{\cal H}_{\rm qu.matter} \}
     \ \Psi = 0 \ .
$$
We will not be more specific about what $\hat{\cal H}_{\rm qu.matter}$
is. We will examine only a simple case where the quantum matter Hamiltonian
has eigenfunctions

$$
   \hat{\cal H}_{\rm qu.matter}\ \Psi_E = EM^2 \ \Psi_E \ ,
$$
with the eigenvalue independent of $a$. (We expect that this will 
illustrate sufficiently well the scales involved in the problem.)
We have written  the eigenenergy in a way that will be convenient
later. Here $E$ has the same dimension as length. We assume the energy of
the quantum matter to be much less than Planck mass, $EM^2 \ll M$. Thus, the
different scales are related as follows:
\be
      EM \ll 1 \ll CM \ .
\label{allscales}
\ee
As in the quantum mechanics examples, we are interested in the phase
behavior of $\Psi$.  The potential energy 
part of $\hat{\cal H}_{gr.+c.m.}$ has a turning point at the maximum
size of the universe. After adding quantum matter, the location of the
turning point ({\em i.e.}, the maximum size of the universe) depends
on $E$. The new classical turning point is at 
$$
    \amax (E) = \sqrt{C^2 + E^2} + E 
$$
and the classically allowed region is $a\leq \amax(E)$. We know that
in the tunneling region the solution can only be exponentially
decaying since the tunneling region is infinite. Thus, from the WKB
joining conditions we know that far away in the classically allowed
region the wavefunction will reduce to a superposition
$$
   \Psi \sim e^{iS+i\frac{\pi}{4}} + e^{-iS-i\frac{\pi}{4}} 
$$
where 
$$
   S = \int^{\amax (E)}_a \ da' p(a') 
$$
with the momentum
$$
  p(a) = M^2\ \sqrt{C^2+2aE-a^2} \ .
$$
Notice that the gravitational `potential' is smooth, and thus the
constant phase shift is just $\pi /4$, independently of the energy
of the quantum matter. Thus this phase shift does not spoil the semiclassical 
approximation. 
The result of the integration is
$$
   S = \frac{(CM)^2}{2} \ (1+\ec^2)\ \{  \frac{\pi}{2} - \sin^{-1} 
      (\frac{\ac-\ec}{\sqrt{1+\ec^2}}) -  (\frac{\ac-\ec}{\sqrt{1+\ec^2}})
      \sqrt{1-(\frac{\ac-\ec}{\sqrt{1+\ec^2}})^2} \ \} \ ,
$$
where we introduced dimensionless variables $\ac = a/C,\ec = E/C$. Notice that
$S$ is proportional to the square of the maximum size $a_{\rm max} (0) = CM$.
Following Section 2.3, we expand $S$ in 
the rescaled\footnote{From (\ref{allscales}) we
see that $\ec \ll 1/CM \ll 1$.} quantum matter energy $\ec$:
\be
   S(\ac ,\ec ) = S (\ac ,0) + \ec \ \frac{\pat S}{\pat \ec} (\ac ,0)
                 + \half \ \ec^2 \ \frac{\pat^2 S}{\pat \ec^2} (\ac ,0)
                 + \cdots \ \ .
\label{expandS}
\ee
The first term is the solution of the classical Hamilton-Jacobi equation 
in the semiclassical approximation. The coefficient of the second
term is 
\be
      \frac{\pat S}{\pat \ec} (\ac ,0) = (CM)^2 \sqrt{1 -\ac^2} \ .
\label{2nd}
\ee
The scale factor $\ac$ is related to the WKB time by $\ac^2 = 1-(t/CM)^2$,
so (\ref{2nd}) becomes
\be
     \frac{\pat S}{\pat \ec} = CM t \ ,
\label{pp}
\ee
as expected.  The coefficient of the third term turns out to be
\be
    \frac{\pat^2 S}{\pat \ec^2} = (MC)^2 \ \{ \frac{\pi}{2} 
    - \sin^{-1} [\sqrt{1-(\frac{t}{CM})^2}] 
    + \sqrt{(\frac{CM}{t})^2 -1} \ \} \ \ .
\label{3rd}
\ee
As Figure 1 shows, away from the vicinity of the 
turning point  (\ref{3rd}) quickly decays to order $(CM)^2$.
\def\fpsangle{0}
\begin{center}
\leavevmode
\fpsxsize 2.5 in
\fpsbox{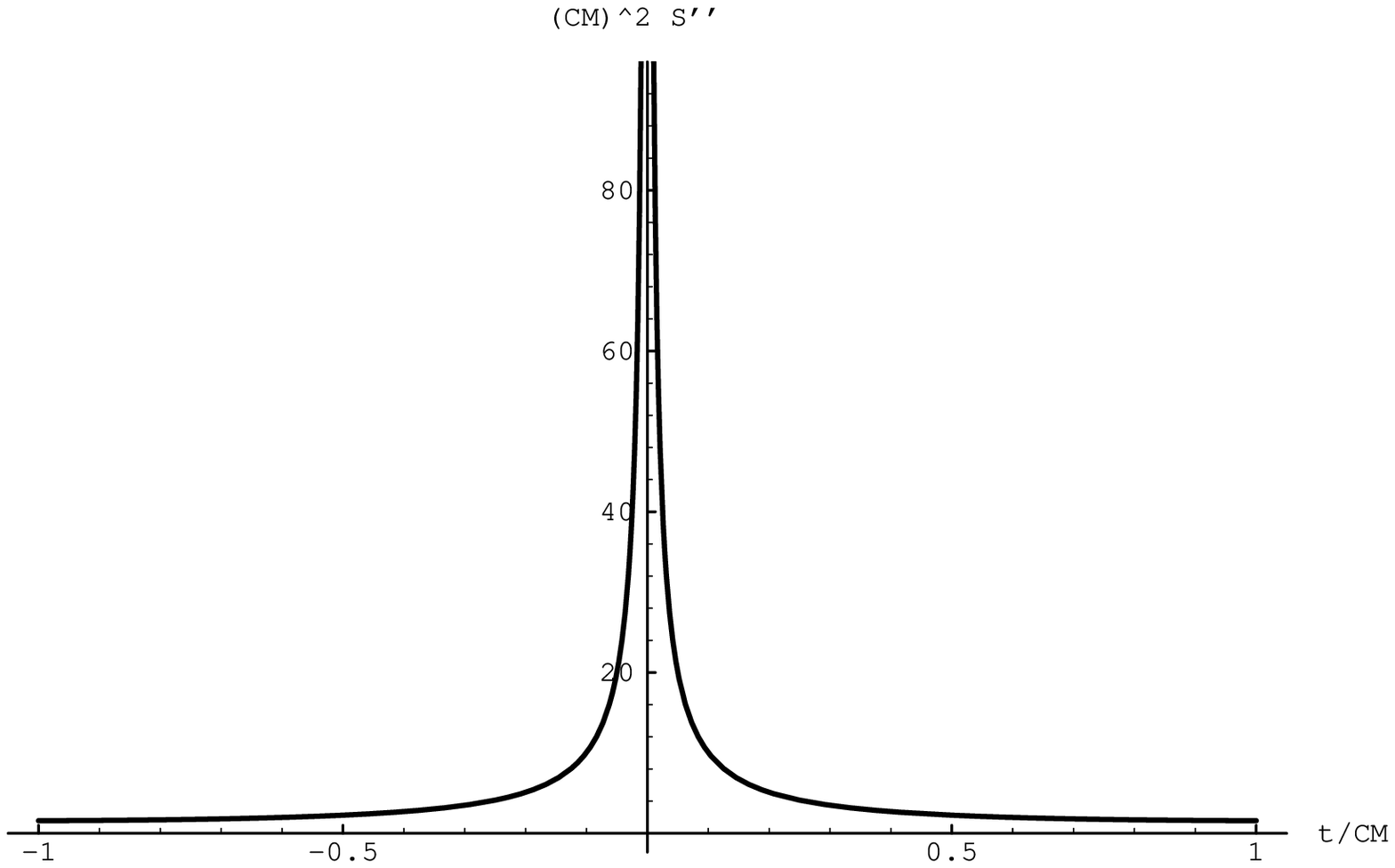}
\end{center}
\begin{quotation}
\small
\noindent
{\bf Figure 1:}
Time dependence of the second derivative of S as a function of $t/CM$.
\end{quotation}

Thus we see that (\ref{pp}) gives a phase to eigenstates proportional to
their energy, so it represents quantum mechanical evolution 
expected from the semiclassical approximation. The 
correction (\ref{3rd}) on the other hand gives
a term proportional to the square of the energy, $\sim (EM)^2$, and so
cannot be absorbed into a quantum mechanical evolution with the matter
Hamiltonian $\hat{\cal H}_{\rm qu.matter}$.

The key fact is that the nonzero end value 
for the third term of (\ref{expandS}) is 
of the order $(EM)^2$  
The correction  is thus negligibly small as
long as the energy of the quantum matter is much less than than Planck
energy (\ref{allscales}). 
If the energy of the quantum matter is larger than the Planck energy, then 
we can still get the correction terms to be small if we consider evolution over
times small compared to the age $CM$ of the Universe; the required 
restrictions can be found from (\ref{3rd}).

\section{Turning Points and Black Hole Evolution in Dilaton Gravity}

Models of 1+1 dimensional gravity have been around for 
quite a while \cite{JT}. More recently, models with
black hole like solutions have become popular.
We are especially interested in the black hole spacetimes
which are vacuum solutions in the CGHS model \cite{cghs}. This model has
recently been studied in the framework 
of Dirac quantization and solutions to the Wheeler--DeWitt equation
have been found \cite{LGK,CJ}. [The equivalence of the solutions
of \cite{LGK} and \cite{CJ} is shown in \cite{Ben}.]
Consider all hypersurfaces
in a black hole spacetime. The Wheeler--DeWitt equation controls 
evolution from one hypersurface to the next one. In such an evolution, 
turning points may appear as we shall show in this section. We will
examine the behavior of the (gravitational) wavefunction in the vicinity
of a turning point. We need to identify the classically allowed region
where the wavefunction is oscillatory. We will also investigate the
possibility of having exponentially decaying (or growing) behavior
in classically disallowed regions. In other words, we study
if tunneling issues can arise.

\subsection{Hamiltonian Formulation of Dilaton Gravities}

We start with a brief review of the results presented in \cite{LGK,GK,GKL}. 
[For additional discussion, see \cite{lmo}.] The action of 
the CGHS model of dilaton gravity is most commonly
presented in the form
\be
 S = \int \ d^2x \sqrt{-g} \ e^{-2\phi} \ [R + 4(\nabla \phi)^2 
                                             + 4\lambda^2] \ .
\label{c33}
\ee
It is possible to rescale the metric, $(g_{\alpha \beta})
= \frac{1}{8}e^{2\phi} \ (\bar{g}_{\alpha \beta})$, so that the kinetic
term disappears from the action. Then
the action becomes 
\be
 S = \int \ d^2x \sqrt{-\bar{g}} \ [\phib \bar{R} -\vb (\phib)] \ ,
\label{c32}
\ee
where
$$
  \phib = e^{-2\phi} \ .
$$ 
For the CGHS model,  $\vb (\phib ) = -\frac{\lambda^2}{2} $. Other models
of dilaton gravity give arise to different forms of the potential $\vb$.
(For some more details, see Appendix D.) The action (\ref{c32})
is the starting point in the Hamiltonian formulation. To proceed, 
the metric is parametrized by introducing 
the lapse and shift functions $N,\nperp$ 
\begin{eqnarray}
  ds^2 = g_{\alpha \beta} dx^{\alpha} dx^{\beta} &=& \frac{1}{8}e^{2\phi}
 \gbar_{\alpha \beta} dx^{\alpha}dx^{\beta} \\ \nonumber
 \mbox{}  &=& \frac{1}{8} e^{2(\phi +\rhob)} \ [-N^2(dx^0)^2 + (dx^1 +\nperp
 dx^{0})^2] \ .
\label{c34}
\end{eqnarray}
Using the parametrization (44) in (\ref{c32}) one can rewrite
the action in a form from which it is easy to read out the canonical
momenta
\begin{eqnarray*}
\Pi_{\phib} &=& \frac{2}{N} (\nperp \rhob' + \nperp' - \dot{\rhob}) \\
\Pi_{\rhob} &=& \frac{2}{N} (-\dot{\phib} + \nperp \phib') \\
\Pi_{N} &=& \Pi_{\nperp} = 0 
\end{eqnarray*}
where $' = \partial / \partial x^1$ and 
$\dot{\ } = \partial / \partial x^0$.  The Hamiltonian turns out
to be a sum of the time reparametrization and spatial diffeomorphism
constraints
$$
  H = \int \ dx^1 \  [N {\cal H} + \nperp {\cal H}_{\perp}] 
$$
where 
\begin{eqnarray*}
   {\cal H} &=& 2\phib'' - 2\phib' \rhob' - \half \Pi_{\rhob}
\Pi_{\phib} + e^{2\rhob} \vb (\phib)  \\
   {\cal H}_{\perp} &=& \rhob' \Pi_{\rhob} + \phib \Pi_{\phib} -
\Pi'_{\rho} \ .
\end{eqnarray*}
The time reparametrization constraint 
${\cal H}=0$ can be written as a Hamilton-Jacobi (H-J) equation
$$
   \half [\ \frac{\pat S}{\pat \phib}\frac{\pat S}{\pat \rhob} 
    -g[\phib , \rhob ] \ ] = 0 
$$
where
$$
     g[\phib ,\rhob ] =  4\phib'' -4\phib' \rhob' + 2e^{2\rhob} \vb
 (\phib ) \ .
$$
It was found \cite{LGK} that a solution to the H-J equation
is given by a functional
\be
    S[\phib ,\rhob ,\Ccal] = \int dx^1 \{ Q_{\Ccal} + \phib' \ln 
   [\frac{2\phib' - Q_{\Ccal}}{2\phib' + Q_{\Ccal}}] \} 
\label{c35}
\ee
which depends on an integration constant $\Ccal$ through the functional
\be
  Q_{\Ccal} = 2\sqrt{(\phib')^2 + (\Ccal + j(\phib ))e^{2\rhob} }
\label{c36}
\ee
where $j(\phib )$ is given by $dj(\phib )/d\phib = \vb (\phib )$. For
the CGHS model, $ j(\phib ) = -\frac{\lambda^2}{2} \phib $.
The constant $\Ccal$ can alternatively be expressed as a
functional of the canonical 
variables $\rhob ,\phib ,\Pi_{\rhob},\Pi_{\phib}$, using
the relations
\bea
  {\Pi}_{\phib} &=& \frac{\delta S}{\delta \phib} = \frac{g}{Q_{\Ccal}} \\
  \label{c38a}
  {\Pi}_{\rhob} &=& \frac{\delta S}{\delta \rhob} = Q_{\Ccal} \ .
  \label{c38b}
\eea
The result is
\be
  \Ccal = e^{-2\rhob} \ (\frac{1}{4} {\Pi}^2_{\rhob} - (\phib')^2)
   - j(\phib ) \ .
\label{c39}
\ee
So far we have not identified the coordinates $x^0,x^1$.
This can be done after gauge fixing. 
Different choices of $N,\nperp$ correspond to different
gauge choices. A common choice is the conformal
gauge $N=1$, $\nperp =0$, so (44) becomes
\be
  ds^2 = e^{2\rho} \ [(dx^0)^2 - (dx^1)^2 ] 
       = \frac{1}{8} e^{2(\phi +\rhob)} \ [(dx^0)^2 - (dx^1)^2 ] \ .
\label{gconf}
\ee
The CGHS model has a global symmetry which allows us to additionally
fix the relation of $\rho$ and $\phi$ to be $\rho = \phi$.
This is called the Kruskal gauge. Now the metric
(\ref{gconf}) can be written as
$$
  ds^2 = -e^{2\phi} \ d\xp d\xm
$$
where $x^{\pm} = x^0 \pm x^1$ are the Kruskal coordinates. 
The gauge choices simplify the expressions of the momenta.
First, in the conformal gauge 
\begin{eqnarray*}
    \Pi_{\phib} &=& -2\dot{\rhob} \\ 
    \Pi_{\rhob} &=& -2\dot{\phib}  \ .
\end{eqnarray*}
Then, in Kruskal gauge $\rhob = \sqrt{8}$ (see (\ref{gconf}), so one of the 
momenta vanishes identically: $\Pi_{\phib} = 0$.

The field equations of the CGHS model have solutions which have the
same properties as static black holes. In the Kruskal gauge, these
are given by    
\be
 e^{-2\rho } = e^{-2\phi} = \frac{M}{\lambda} - \lxp \lxm 
\label{c40}
\ee
where $M$ is the mass of the black hole. We can use this to identify
the integration constant $\Ccal$. First we
express $Q_{\Ccal}$ and $\Pi_{\rhob}$ in the CGHS fields:
\be
  Q_{\Ccal} = 4 \ \sqrt{e^{-4\phi}(\phi')^2 +(2\Ccal  -\lambda^2 e^{-2\phi}) }
\label{c41}
\ee
and
$$
   \Pi_{\rhob} = 4e^{-2\phi} \dot{\phi} \ ,
$$
Then, from (\ref{c40}) we
find $\phi',\dot{\phi}$ and substitute these to
$Q_{\Ccal},\Pi_{\rhob}$. Using (\ref{c38b}) we can finally 
identify\footnote{This formula depends on the choice of the 
integration constant mentioned
in a footnote in Appendix D. 
However, the important point is that after $M$ has been
substituted to the expression for $Q$, the resulting expression contains no 
ambiguous integration constants.}   
\be
       \Ccal = \frac{M\lambda}{2} \ .
\label{c42}
\ee
{}(From now on, we will set $\lambda=1$.)

To conclude this subsection, we briefly mention 
some results from efforts to quantize the
theory.
Following the Dirac quantization approach in
the Schr\"{o}dinger representation, the Hamiltonian constraint is replaced
by the Wheeler--DeWitt equation. 
In \cite{LGK} a specific operator ordering was used and the WDW
equation was written  as 
\be
   \hat{\cal H} \ \Psi (\phib ,\rhob ) =
    \half (g - Q_{\Ccal} \ \hat{\Pi}_{\phib} \ Q^{-1}_{\Ccal}
   \ \hat{\Pi}_{\rhob}) \ \Psi (\phib ,\rhob ) = 0 \ .
\label{c37}
\ee
This equation is solved by a wavefunctional $\Psi = \exp (\pm iS)$
where $S$ is the solution (\ref{c35}) of the H-J equation. It is
curious that the exact solution should take the form of a WKB solution.
This seems to be tied with the specific choice of operator
ordering. As we shall discuss, $Q_{\Ccal}$ tends to zero at a
turning point, so (\ref{c37}) becomes singular. 
This suggests that the correct quantization would have
significantly different choice of operator ordering near the turning point,
and such a WKB solution would not be exact\footnote{We should emphasize
that these statements apply to the formulation of dilaton
gravity in the variables and foliations considered in this paper. Although
in the present setting we find turning points and expect the breakdown
of WKB approximation in their vicinity, it does not exclude the possibility
of finding a canonical transformation into some other coordinates in which
the turning point issues do not arise. For instance, in a ``laboratory frame''
the motion of a simple harmonic oscillator has two turning points, and
the WKB approximation would break down in their vicinity. However, after
a canonical transformation to action-angle variables, the turning points 
disappear and the WKB approximation is exact. It may be interesting to
note there are alternative variables for dilaton gravity \cite{cjz} in
which the WKB approximation is again thought to be exact \cite{extra}.
It would be interesting to investigate what 
happens to turning points in the context of \cite{extra}.}.  We will 
thus use the WKB solution in the same spirit as in the 
previous section to study
the potential troubles with the semiclassical approximation 
near the turning point.

\subsection{Turning Points in Black Hole Evolution}

Let us investigate the nature of spacelike hypersurfaces that  we must
consider. By the diffeomorphism constraint we know that the amplitude
for any hypersurface (in the quantum gravity wavefunctional) must depend
only on the intrinsic geometry of the hypersurface, and not
on the coordinates used to parameterize it. To describe this 
intrinsic geometry we consider the scalar function $\phi$ on the hypersurface.
Let $s$ be the invariant length along the 
hypersurface from some fixed point $\phi=\phi_0$, measured in the 
direction of increasing $\phi$, lets say. 
The function $\phi(s)$ gives an intrinsic characterization of the hypersurface.

As an example, consider (in the above black hole geometry) a 
class of hypersurfaces $\Sigma_{\gamma}$ which are straight lines 
in Kruskal coordinates
\be
   \Sigma_{\gamma} \ : \ \ \ \xm = -\al^2 \xp + \gamma \ .
\label{lines}
\ee
For a fixed slope $\al^2$, there is one free
parameter $\gamma $ which can be positive or negative.
The special case
$\gamma =0$  corresponds to constant Schwarzschild time slices
$t = -\ln \al $. For $\Sigma_{\gamma}$ we find 
\be
  |\frac{d\phi}{ds} (\phi) | = \frac{e^{\phi}}{2} 
    \sqrt{(\gamma / \alpha)^2 + 4(e^{-2\phi}-M)} \ .
\label{dpds}
\ee
The equation could be integrated to give $\phi (s)$. Note that 
either sign of $\gamma$ gives the same function $\phi(s)$.  Thus 
each one-geometry
has two embeddings in the spacetime, except
for  $\gamma =0$, where
the one-geometry has only one kind of an embedding.
Note that for a given value of $\phi$ the slope $|\frac{d\phi}{ds} (\phi) |$ is
smallest when the direction of the hypersurface through that point 
corresponds to $\gamma=0$.

The above situation is generic. In any spacetime, pick a point P
(with value of scalar $\phi=\phi_0$), and consider all possible 
infinitesimal spacelike line segments through this point. The quantity 
 $|\frac{d\phi}{ds} (\phi_0) |$ reaches a  minimum on  a certain 
unique direction of the line segment; call this segment $\sigma_0$, and let
the value of  $|\frac{d\phi}{ds} (\phi_0) |$ on $\sigma_0$ be 
called  $|\frac{d\phi}{ds} (\phi_0) |_{\rm min}$. This line segment 
partitions the set of 
all spacelike segments at our point into two connected classes.  
For any given value of $|\frac{d\phi}{ds} (\phi_0 ) |$ at our point, with
\be
|\frac{d\phi}{ds} (\phi_0) |_{\rm min} ~< 
|\frac{d\phi}{ds} (\phi_0) | ~<\infty \ ,
\label{dpdsp}
\ee
there is exactly one line segment in each of the two classes, call these
segments as $\sigma_{\pm}$, 
matching with that value of $|\frac{d\phi}{ds} (\phi_0) |$. We label these 
two classes as {\it category I} and {\it category II} as shown in Figure 2.

\def\fpsangle{270}
\medskip
\begin{center}
\leavevmode
\fpsxsize 2 in
\fpsbox{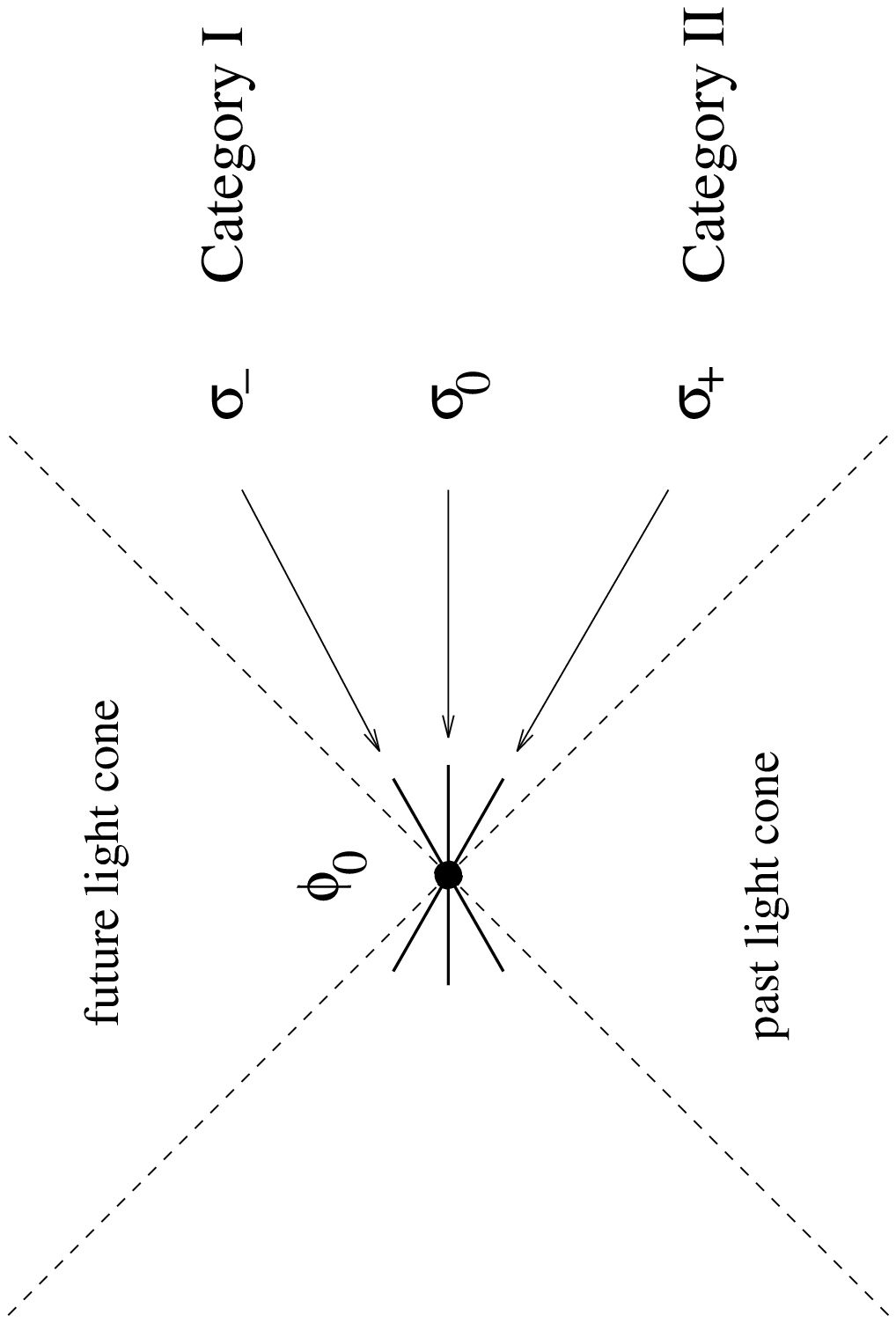}
\end{center}
\vskip 0.5 true cm
\begin{quotation}
\small
\noindent
{\bf Figure 2:} Local definition of the two classes of segments.
\end{quotation}
\medskip
Since the `special direction' given by the segment $\sigma_+$ at
each spacetime point is continuous as we move the point P 
(as is the  direction  in the other category given 
by $\sigma_-$), we can use the above split of the 
tangent space at each point to define two categories of hypersurfaces.
A category I hypersurface has its tangent direction at each 
point in category I of tangent directions, and similarly for 
category II. (We ignore for the discussion below the case where 
the hypersurface alternates between categories I and II.)

Thus in the above example of the black hole geometry the 
surfaces $\Sigma_{\gamma}$ with $\gamma <0$ belong to category I
and the with $\gamma >0$ belong to category II. The constant 
Schwarzschild time hypersurfaces lie at the intersection of the two categories.

Now we note that the hypersurfaces at the boundary of categories I and II
are turning points in the evolution given by the Wheeler--DeWitt equation.
In the simple 1-dimensional problems considered in section 2 we 
see that the classical trajectory passes twice through each point $x_1$, once 
with positive momentum (going towards the turning point) and once 
with negative momentum (receding from the turning point). At the turning
point, the momentum is zero. 
The analogue of the coordinate $x_1$ is the `gravitational variable', the
1-geometry, and as we saw above, at least 
locally the same intrinsic 1-geometry 
embeds in two different ways in the classical 
spacetime, once in category I and once in category II.  We will 
show now that the momentum  of the 
gravitational field tends to zero at the boundary 
of the two categories. Further, the absolute value of the
momentum is the same when we consider the same 
intrinsic 1-geometry in the two categories I and II. This is analogous to
the particle quantum  mechanics case where
the absolute value of the momentum is the same at point $x_1$ both on
the trajectory leading towards  the turning point and on the trajectory 
returning from the turning point. 

For simplicity 
let us consider again the surfaces $\Sigma_{\gamma}$ of (\ref{lines}) (in
the case of an eternal black hole). 
Let us rewrite the defining equation as 
\be
      x^0 = \frac{1-\al^2}{1+\al^2} x^1 + \frac{\gamma}{1+\al^2} \ .
\label{skewline}
\ee
If $\al^2 =1$, these are just constant Kruskal time hypersurfaces
$x^0 = \gamma /2$.
In the case $\al^2 \neq 1$ it is convenient to use a boosted
Kruskal frame\footnote{Boosts do not affect 
the Kruskal gauge $\rho = \phi$.} $\tilde{x}^{\pm}$ with
\begin{eqnarray}
  \tilde{x}^0 &=& x^0 \ \cosh \Theta  - x^1 \ \sinh \Theta  \\ \nonumber 
  \tilde{x}^1 &=& -x^0 \ \sinh \Theta +  x^1 \ \cosh \Theta \ . 
\label{boost}
\end{eqnarray}
with a boost angle $\Theta$ given by
$$
    \tanh \Theta \equiv \frac{1-\al^2}{1+\al^2} \ .
$$
In the new frame, (\ref{skewline}) is seen to 
correspond to constant time slices
$\tilde{x}^0 = \tilde{\gamma} /2$ with $\tilde{\gamma} = \gamma / \al$.  
Now 
$$
    e^{-2\phi} = M - \frac{\tilde{\gamma}^2}{4} + (\tilde{x}^1)^2 
$$
so
\be
   \phi'  \equiv \frac{d\phi}{d\tilde{x}^1}(\phi) = -e^{2\phi } \ \tilde{x}^1 
\label{c43}
\ee
for a $\Sigma_{\gamma}$ hypersurface. 
We already found what $Q_{\Ccal}$ looks
like in the CGHS notation, and similarly 
$$
  g = 8e^{-2\phi} \ [2(\phi')^2 -\phi'' - e^{2\rho}] \ .
$$
Substituting (\ref{c43}) to $Q_{\Ccal}$ and $g$ yields
\bea
         \Pi_{\rhob} &=& Q_{\Ccal} = 2 |\tilde{\gamma} | \\ \nonumber
         \Pi_{\phib} &=& \frac{g}{Q_{\Ccal}} \equiv 0 \ .
\eea
The latter equation is just what we expect from our previous observation that
$\Pi_{\phib} \equiv 0$ in Kruskal gauge.  
More important is
that the first equation 
shows that as $\gamma$ changes from negative to positive
(crossing the boundary of categories), the momentum $\Pi_{\bar{\rho}}$
decreases to zero at the boundary and then increases again; the absolute
value is the same for  corresponding points in the two 
categories; i.e., surfaces described by $\pm \gamma$. This is
consistent with the
boundary of the two categories being a {\em turning point in dilaton gravity}.
(As noted above, the turning point hypersurfaces are constant 
Schwarzschild time slices.) 

\subsection{Absence of Tunneling}

Recall the situation in our earlier
1-dimensional quantum mechanical examples. The classical trajectory 
reflects off the turning point. The quantum wavefunction decays 
exponentially inside the barrier, and we had assumed that this barrier 
grows without bound as we go towards large $x$, so that there is no 
tunneling. This was also the case
in the minisuperspace closed cosmology. 
Let us verify that an analogous situation holds in the 
dilaton gravity case too.

We illustrate in Figure 3 below the situation  
for a particle moving in a linear 
potential $U(x) = x$. The momentum $p$ becomes imaginary in 
the region $x >E$. In the dilaton gravity case, we consider a 1-parameter 
family of 1-geometries foliating the spacetime; this parameter will be 
analogous to the coordinate $x$. 
The energy $E$ of the quantum mechanical 
example will turn out to be analogous to the mass $M$ in the black 
hole spacetime. We will see that 
the wavefunctional gets increasingly suppressed as we 
move deeper into the classically disallowed region along the parameter 
labeling the 1-geometries, so there is no hint of a 
finite tunneling direction in this simple model.

Let us examine how the gravitational
wavefunction behaves near the turning
point. We rewrite $S$ in a geometric form
$$
   S = \int \ ds \ e^{-2\phi}\{ 4q -2\frac{d\phi}{ds} 
 \ln [\frac{d\phi /ds + q}
    {d\phi /ds - q}] \}
$$
where $s$ is the invariant length along a one-geometry and
$$ 
    q \equiv q [\phi (s)] = \sqrt{(\frac{d\phi}{ds})^2
     + (Me^{2\phi} -1)} \ .
$$
With this notation, the 
wavefunctional $\Psi = \exp{iS}$ explicitly depends 
on a one-geometry $\phi (s)$.
At a turning point, the exponent $S$ is zero. 

A possibility to consider is that at the turning point the
one-geometries might tunnel into the direction where they can
only fit into spacetimes of more massive black holes. 
For example, consider
the direction of one-geometries with 
$$
 \dpds (\phi) = \sqrt{1 - (\bar{M}-\frac{\gamma^2}{4})e^{2\phi}}
$$ 
where $\bar{M} > M$. These can fit in the spacetime of a higher 
mass $\bar{M}$ black
hole as slices $x^0 = \gamma^2 /4$.  
They give arise to a momentum
\begin{equation}
  \Pi_{\rhob} = Q_M = 4 \sqrt{M-(\bar{M}-\frac{\gamma^2}{4})}
\label{impi}
\end{equation}
which becomes imaginary if $\bar{M}-\gamma^2 /4 > M$. Let us check how the
wavefunctional behaves in this case. Rewriting $Q_M$ as $Q_M = i2\delta$,
the exponent of the wavefunctional is
\begin{eqnarray*}
 S &=& \int^{\infty}_{-\infty} dx^1 \ \{ Q_M + \bar{\phi}' 
       \ln [\frac{2\bar{\phi}' - Q_M}{2\bar{\phi}' + Q_M} ] \} \\
   &=& i\int^{\infty}_{-\infty} du \ \{ \delta 
                             - \tan^{-1} (\frac{u}{\delta}) \} 
             = \frac{i\pi \delta^2 }{2} \ .
\end{eqnarray*}
After resubstituting $\delta$ we find that the wavefunctional behaves
as
\be
   \Psi = \exp \{ -2\pi [\bar{M}-\frac{\gamma^2}{4} -M] \} \ \ .
\ee
We interpret the situation as follows. Let us compare (\ref{impi})
with a momentum of a particle in a one-dimensional potential, 
$p = \sqrt{2m(E-U(x))}$. We can identify $M$ in (\ref{impi}) as the total
gravitational (ADM) energy, which is the analogue of the total energy
$E$ in the particle mechanics case. Similarly we can identify 
$\bar{M} - \gamma^2 /4$ as a coordinate, analogous to $x$. 
Thus the situation is analogous to a particle moving in a linear 
potential $U(x) = x$, where the momentum $p$ becomes imaginary in 
the region $x >E$. We illustrate this in Figure 3.

\medskip
\begin{center}
\leavevmode
\fpsxsize 1.5in
\fpsbox{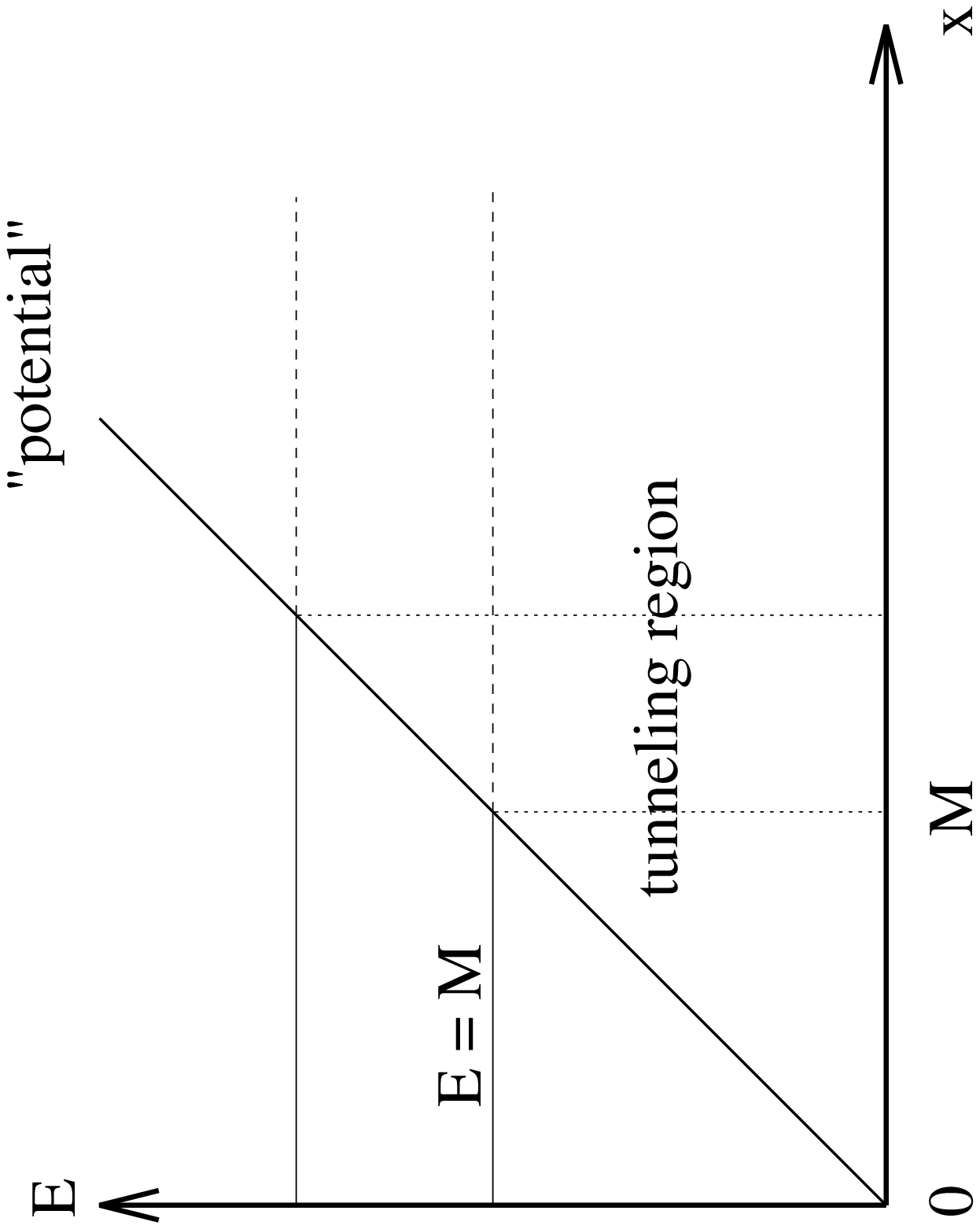}
\end{center}
\vskip 0.5 true cm
\begin{quotation}
\small
\noindent
{\bf Figure 3:}
The classically allowed region and the tunneling region.
\end{quotation}
\medskip
 
The tunneling direction is infinite, so the one-geometries cannot 
escape from the classically 
allowed region. Further, the gravitational `potential'
appears to be smooth. Therefore in this case
we cannot see any hint of
a behavior which would be required for the deviations from semiclassical
approximation which were discussed in Section 2.

\subsection{Summary}

We have seen in this section that the 
gravity part of the wavefunctional has turning points in the 
evolution given by the Wheeler--DeWitt equation, and that 
there are no finite tunneling directions from the turning point.
A consequence of these turning points is that the semiclassical
evolution \cite{dealwis, demkief} of matter breaks 
down in the vicinity of the turning point.
For example, in \cite{demkief} it was shown how the semiclassical
expansion yields a Tomonaga--Schwinger equation
\be
   \frac{-i}{2} ( \frac{\delta S}{\delta \phib} \frac{\delta }{\delta \rhob}
     + \frac{\delta S}{\delta \rhob} \frac{\delta}{\delta \phib} ) \chi
 = {\cal H}_m \chi
\label{TSeqn}
\ee
for the evolution of the matter wavefunctional $\chi$.
S is the solution
of the H-J equation, and ${\cal H}_m$ is the Hamiltonian density of the
matter field $f$ in the Schr\"{o}dinger picture.\footnote{Note that
our conventions are opposite of those of \cite{demkief}: we denote
the dilaton and the conformal factor which appear in the
action (\ref{c32}) as $\phib ,\rhob$ whereas in \cite{demkief} a notation
$\phi ,\rho$ is used.}

  If there are no turning points, this equation can
be formally integrated to yield the functional Schr\"{o}dinger equation for
the free scalar field on the classical background spacetime. However, 
there is an obstacle if there are turning points. This equation 
cannot evolve the matter wavefunctional from slices of category I to 
slices of category II. This is because if we write the 
wavefunctional for matter as a function of 
variables that give the intrinsic 1-geometry of a slice, then 
both before and after the turning point we encounter the 
same 1-geometry. We would like the
matter to evolve to different values on these two different 
slices with the same intrinsic geometry, but a first order 
evolution equation does not permit that. In handling the full 
second order Wheeler--DeWitt equation we have
to understand the physics of the turning point. But from the 
discussion above (in particular the absence of tunneling directions) we 
expect that the quantum mechanical 
examples of the earlier sections should be a 
reliable guide. Thus we would expect that while the
matter wavefunction gets distorted away from its 
semiclassical form near the turning point\footnote{It would be interesting
to estimate the ``size'' of the breakdown region, in the manner
of Section 2. For example, for black holes the size of the region
could depend strongly on the time of the turning point constant
Schwarzschild time slice. This is suggested by the large distortion of
a matter wavefunction from its semiclassical form on 
the S-slices of \cite{klmo}.}, 
it regains its 
semiclassical form away from the turning point. 
Thus we can patch together the solution of the 
Wheeler--DeWitt equation in the two categories, and 
have a complete solution on all hypersurfaces. 

\section{Discussion}

In this paper we have examined the effect on the evolution of 
matter modes when the gravity wavefunctional passes through a 
turning point. At such a point the usual decomposition
of the complete wavefunctional, into a `fast' gravity part 
giving a classical spacetime and a `slow' part giving matter 
evolution on spacetime, breaks down.One then asks what effect this 
apparent breakdown of the semiclassical approximation creates on the 
matter wavefunction after we recede from the turning point. As
we saw, there are two cases to analyze. In one case the `time' for
which the gravity mode ceases to be `fast' is so small that the matter 
wavefunction does not evolve significantly in that time; here it is 
easy to conclude that there is no lasting effect
of the turning point. In the other case the matter wavefunction does 
evolve significantly in this period, and a more detailed analysis  
shows that there is again no lasting effect after we recede from the 
turning point, provided that there are no tunneling directions available 
at the turning point\footnote{If we make a canonical transformation on the 
position and momentum of the heavy particle, then the turning point in the
new coordinate occurs at a different point in the heavy particle trajectory.
In cases where we can make such a canonical transformation, we can understand
why there should be no large effects on the light particle wavefunction due to
the occurence of a turning point in the heavy particle motion.}. 
(If there are tunneling
directions then we cannot approximate the wavefunction as being composed 
only of the incident and reflected parts at the
turning point.)

The above issue will arise in any approach to quantum gravity that
uses a wavefunctional formalism. How do we fit together parts of
the solution to the Wheeler--DeWitt equation that are found in the
different domains where semiclassical evolution is
well defined? At the boundaries of such regions in superspace
we must carefully check the `joining conditions' to continue
the wavefunctional, in particular noting any 
allowed tunneling directions. 

In the black hole context this issue is relevant because
the Schwarzschild coordinate system determines a slicing
that approaches a turning point of the second kind described above.
This leads to an apparent violation of the semiclassical wavefunctional 
near the horizon, and gives apparent large 
effects from gravity fluctuations in calculations in any expansion scheme
that uses the Schwarzschild coordinate system. Such effects must be
carefully removed before looking for possible violations
of standard physics. Note that our considerations involved
only matter and gravity having simple local actions; we have
thus not examined the consequences of the existence of
extended strings or D-branes.

\bigskip

\bigskip

{\Large \bf Acknowledgements }

\bigskip

We would like to thank S. Coleman, A. Guth, and J. Preskill for discussions
at various stages of this investigation.

\bigskip

\bigskip

\Large
\appendix{{\bf Appendix A}}
\normalsize

\bigskip

Consider the Schr\"{o}dinger equation (\ref{SE}) with the
Hamiltonian (\ref{hamilton}): 
\be
      \hat{H} =
       -\frac{\hbar^2}{2M} \frac{\pat^2}{\pat X^2} + m_1 rX 
     -\frac{\hbar^2}{2\mu} \frac{\pat^2}{\pat y^2} +\frac{\kappa}{2}
     (y - \frac{r\mu}{\kappa})^2  \ .
\label{redundant}
\ee
We now use the semiclassical approximation to find an approximate
solution $\Psi_{sc}$ of (\ref{SE}), in a manner 
in which it is most convenient to compare with the exact solution. Thus, 
we will keep using the coordinates $X,y$, and use the total mass
$M$ as the expansion parameter in the semiclassical approximation.
So we will find the approximate wavefunction $\Psi_{sc} = \exp 
\{ \frac{i}{\hbar} [ \scal_0(X) + M\scal_1 (X,y) ] \}$. We want the classical
motion of $X$ to be independent of the mass $M$, so that it defines a good
clock as we discussed in Section 2.1. So we rewrite the Schr\"{o}dinger
equation (\ref{SE}) in the form
\be 
      \{ \hat{H}_X +\hat{H}_{Xy} \} \ \Psi =
      \left\{ 
      \left[ -\frac{\hbar^2}{2M} \frac{\pat^2}{\pat X^2} 
              + MrX - E_{\rm total} \right]
     + \left[ -\frac{\hbar^2}{2\mu} \frac{\pat^2}{\pat y^2} 
       +\frac{\kappa}{2} (y - \frac{r\mu}{\kappa})^2 - m_2rX \right]
     \right\} \ \Psi = 0 \ .
\ee
The leading order approximate solution is then
\be
         \Psi_{sc} (X,y) \approx \frac{1}{(-2V(X))^{\frac{1}{4}}}
              \ e^{\pm \frac{i}{\hbar} M \scal_0(X)} \ \chi (X,y) \ ,
\ee
where
$$
\left\{
\begin{array}{l}
       \scal_0(X) = \frac{2}{3} \sqrt{2r} [a(0) - X]^{\frac{3}{2}} \\
       a(0) \equiv E_{\rm total}/(Mr) \\
       V(X) = r(X-a(0)) \ \ \ .
\end{array} \right.
$$ 
The function $\chi$ satisfies 
\be
   i\hbar \frac{\pat \scal_0}{\pat X} \frac{\pat}{\pat X} \ \chi =
\hat{H}_{Xy} \ \chi \ .
\label{c28}
\ee
The WKB time is defined using the classical trajectory of the center of
mass:  
\be
   t_{\rm WKB} (X) = \int^a_X \frac{dX'}{\pat \scal_0 /\pat X} 
               = \pm \sqrt{\frac{2}{r}} \ [a(0)-X]^{\half} \ ,
\label{c29}
\ee
note that in the leading order semiclassical approximation
the location of the turning point $a$ is independent of the
energy of the light particle. The overall sign choice will be chosen
to give a negative (positive) $t_{\rm WKB}$ for a motion towards (away from) 
the turning point. 

Next, we expand $\chi$ in the eigenenergy $E_n$
eigenfunctions $G_n(y)$ of the simple harmonic oscillator; these satisfy
$$
    \hat{H}_{Xy} \ G_n(y) = (E_n -m_2rX) \ G_n(y) \ .
$$ 
According to  (\ref{c28}),
(\ref{c29}) $\chi$ then time evolves as
$$
   \chi = \sum_n c_n(t_{\rm WKB}) \chi_n 
      = \sum_n c_n(t_{\rm WKB,0}) e^{-\frac{i}{\hbar} E_n 
         (t_{\rm WKB} - t_{\rm WKB,0}) + \frac{i}{\hbar}
      m_2 r \int^{t_{\rm WKB}}_{t_{\rm WKB,0}} \  X(t')dt'} \ G_n(y) \ .
$$
The second term in the exponent is independent of the energy levels and
can be dropped. However, in general one needs to be more careful.
A time evolution where $X$ propagates to the
turning point and back to the same point again is periodic. This may
give arise to an additional Berry's phase 
in general (multidimensional) models. 

\bigskip

\bigskip

\Large
\appendix{{\bf Appendix B}}
\normalsize

\bigskip

Consider the same model as in Section 2, but change the potential
$V(X)$ to be
$$
  V(X) = \left\{ \begin{array}{l} 0 \ , \ X \le 0 \\
                                  V_0 >0\ , \ 0 \le X  
                                 \end{array} \right.
$$
 Now the equation (\ref{mark}) becomes
$$
  [-\frac{\hbar^2}{2M} \frac{\pat^2}{\pat X^2} + m_1V(X)] \ F_n(X) 
  = (E_{\rm total}-E_n) \ F_n(X) \ .
$$
In  the region $X \le 0$, the exact wavefunction
of the system is then
$$
   \Psi = \Sigma_n \ c_n \cos [k_n X + \phi(E_n)] G_n (y) 
$$
where
$$
   k_n = \frac{1}{\hbar} \sqrt{2M(E_{\rm total}-E_n)} \ \ .
$$
We will assume that the energy of the heavy 
particle  $E_{\rm total} -E_n$ is less than $V_0$; thus 
classically the heavy particle reflects off the potential 
barrier at the same position $X=0$ for all energies  $E_n$ of 
the light particle. Thus there is no effect of the form (\ref{c211b}) 
which comes from a nonlinear change in the reflection position 
with changing  $E_n$. But a simple calculation shows  that 
$$
     \phi (E_n) = \tan^{-1} (\sqrt{\frac{V_0 - (E_{\rm total}-E_n)}
          {E_{\rm total}-E_n}}) \ .
$$
Thus, $\phi (E_n)$ 
ranges from $\pi /2$ to 0 as $E_{\rm total}-E_n$ ranges from 0 to $V_0$; the 
phase shift is zero when the turning point
is just below the edge of the potential step. 
This illustrates that the phase shift (\ref{pshift}) can indeed depend
on the location of the turning point. We now get 
for the light particle wavefunction the 
relations (\ref{c211bpone}), (\ref{c211bptwo}) with
\be
   c_n^{\rm after}~=~ c^{\rm before}_n e^{-\frac{i}
                             {\hbar}E_n \Delta t_{\rm WKB}} e^{2i\phi(E_n)}   
\label{c211bpa}
\ee

\bigskip

\Large 
\appendix{{\bf Appendix C}}
\normalsize

\bigskip

In this appendix, we give an argument why the $\gamma$-dependent term
in the minisuperspace model (\ref{minisuper}) of Section 3 can be 
dropped in the WKB regime.
First, let us rewrite the Wheeler--DeWitt equation (\ref{minisuper}) using 
dimensionless
variables and parameters. We rescale the scale factor $a$ and the maximum
size of the universe $C$ as
$$
   a \mapsto \tilde{a} \equiv Ma \ ,  \quad C \mapsto \tilde{C} \equiv MC \ .
$$
In other words, $\tilde{a}$ and $\tilde{C}$ are given in Planck units.
Dropping the tildes, the rescaled WDW equation reads
$$
  \half \{ \frac{\pat^2}{\pat a^2} + \frac{\gamma}{a} \frac{\pat}{\pat a}
       + (C^2 - a^2 ) \} \Psi (a) = 0 .
$$
The WKB form for the wavefunctional $\Psi$ is 
$$
        \Psi \approx \exp \{i\int^C_a da' p(a') \}
$$
where $p(a)$ is the momentum
$$
       p(a) = \sqrt{C^2 - a^2} \ .
$$
Substituting this ansatz to the WDW equation, we find that the
kinetic terms $\pat^2 / \pat a^2 + (\gamma / a) \pat / \pat a$ 
give a contribution
\be 
    -ip' - p^2 - i \frac{\gamma }{a} p 
\label{everything}
\ee
where $p' = dp/da = -a/p$. We cannot use the WDW approximation if the
universe is of the order of Planck scale or smaller. So we must consider
the region $a \gg 1$. Further, the validity of the WDW approximation 
requires that
$$
     p^2  \gg |p' | \ . 
$$
In this case, we can drop the $p'$ term in (\ref{everything}). From the
above two requirements we find that
$$
   p^3  \gg a \gg 1
$$
which also implies that
$$
   p^2  \gg p \gg \frac{\gamma}{a} p  \ .
$$
This means that we can also drop the $\gamma$-dependent term from the
kinetic term of the WDW equation.

\bigskip

\Large
\appendix{{\bf Appendix D}}
\normalsize

\bigskip

Consider a class of two dimensional dilaton
gravity models of the form
\be
  S = \int \ d^2x \sqrt{-g} [\half g^{\alpha \beta} \pat_{\alpha} \chi 
     \pat_{\beta} \chi - V(\chi) + D(\chi) R] \ .
\label{AppC1}
\ee
where $\chi$ is the dilaton scalar field and $(g_{\alpha \beta})$ is a
two dimensional metric. To rewrite the action in a form where
the kinetic term disappears, one first needs to rescale the metric
$$
  (g_{\alpha \beta}) = \Omega^{-2} (\bar{g}_{\alpha \beta}) 
$$
with a rescaling factor $\Omega$ satisfying the equation
$$
    -1 + 4 \frac{dD}{d\chi} \frac{d\ln \Omega}{d\chi} = 0\ ,
$$
then perform a field redefinition
$$
   \chi \rarr \bar{\phi} = D(\chi) 
$$
and introduce a new potential
$$
  \bar{V}(\bar{\phi}) =
\frac{V(\chi (\bar{\phi}))}{\Omega^2(\chi(\bar{\phi}))} \ .
$$
As a result, the action (\ref{AppC1}) becomes 
\be
 S = \int \ d^2x \sqrt{-\bar{g}} \ [\phib \bar{R} -\vb (\phib)] \ .
\label{AppC2}
\ee
In the case of the CGHS model, the action is most commonly
presented in the form
\be
 S = \int \ d^2x \sqrt{-g} \ e^{-2\phi} \ [R + 4(\nabla \phi)^2 
                                             + 4\lambda^2] \ .
\label{AppC3}
\ee
This form of the action can be 
related to (\ref{AppC1}) and (\ref{AppC2}) with
the identifications\footnote{Note that $\Omega$ and $\bar{V}$ are 
defined up to an
arbitrary integration constant, we have set it equal to 1.}
\begin{eqnarray*}
  \phib &=& D(\chi) = \frac{1}{8} \chi^2 = e^{-2\phi} \\ 
 \Omega^2 (\chi) &=& \chi^2 \\
 \vb (\phib ) &=& -\frac{\lambda^2}{2} \ .
\end{eqnarray*}










\end{document}